\documentclass[11pt]{article}

\input epsf
\usepackage{graphicx}
\usepackage{amsmath}
\usepackage[ruled,linesnumbered]{algorithm2e}
\usepackage[numbers]{natbib}
\usepackage[margin=1.0in]{geometry}
\usepackage{hyperref}
\usepackage{todonotes}
\usepackage{authblk}

\newcommand{\enc}[1]{[\![{#1}]\!]}
\newcommand{\specialcell}[2][c]{\begin{tabular}[#1]{@{}c@{}}#2\end{tabular}}
\begin{document}

\title{\LARGE SEEK: model extraction attack against hybrid secure inference protocols}
\author[1]{Si Chen\thanks{si.chen@osr-tech.com}}
\author[2]{Junfeng Fan\thanks{fan@osr-tech.com}}
\affil[1,2]{Open Security Research, Shenzhen, China}
\date{\vspace{-5ex}}

\maketitle

\begin{abstract}
Security concerns about a machine learning model used in a prediction-as-a-service include the privacy of the model, the query and the result. Secure inference solutions based on homomorphic encryption (HE) and/or multiparty computation (MPC) have been developed to protect all the sensitive information. One of the most efficient type of solution utilizes HE for linear layers, and MPC for non-linear layers. However, for such hybrid protocols with semi-honest security, an adversary can malleate the intermediate features in the inference process, and extract model information more effectively than methods against inference service in plaintext. In this paper, we propose SEEK, a general extraction method for hybrid secure inference services outputing only class labels. This method can extract each layer of the target model independently, and is not affected by the depth of the model. For ResNet-18, SEEK can extract a parameter with less than 50 queries on average, with average error less than $0.03\%$.
\end{abstract}


\section{Introduction}

For a machine learning model used in a prediction-as-a-service (PaaS) setting, the model provider usually is concerned about the privacy of the deployed model. Revealing the model information will enable a user to develop his own model. In addition, the model information can be reverse-engineered to reveal its training data \cite{fredrikson2015model, shokri2017membership}, or enable an attacker to fabricate adversarial samples \cite{lowd2005adversarial}. On the other hand, users of PaaS may have privacy concerns about the input data, and do not want to upload the input in plaintext to a service hosted by the model provider. Thus neither the server side nor the client side is a satisfactory place to perform the model inference computation.

To solve this dilemma, secure inference protocols are proposed, which enables the client to query a model deployed in a remote server, while preventing the client and server to learn any additional information. Secure inference solutions are based on homomorphic encryption (HE), multiparty computation (MPC), or both families of techniques.
Solutions based on homomorphic encryption suffer from limitation of the practical FHE schemes. The levelled, and relatively efficient FHE schemes, including BFV, BGV, and CKKS, support fixed number of multiplications without bootstrapping. 
By replacing the activation functions with polynomial functions, the levelled FHE schemes can compute both the linear layers and non-linear layers, but cannot support multiplication depth needed by a deep neural network with more than 3 or 4 layers \cite{gilad2016cryptonets, brutzkus2019low}. On the other hand, the most efficient secure inference protocol based on MPC either use garbled circuit and generally incur higher communication cost \cite{rouhani2018deepsecure,ball2019garbled,riazi2019xonn}, or require three non-colluding parties \cite{mohassel2018aby3,wagh2019securenn,kumar2020cryptflow}, which is an additional requirement not readily satisfied in practice.

To perform secure inference for deep neural networks, while utilizing the efficiency of levelled FHE schemes, hybrid solutions based on FHE and MPC emerged \cite{juvekar2018gazelle,boemer2019ngraph,mishra2020delphi,rathee2020cryptflow2,huang2022cheetah}. The linear part of a neural network, which contains the majority of computation cost, is processed by a FHE scheme, while the non-linear part is process by a MPC scheme. Between linear layers and non-linear layers, a pair of protocols are performed to transfer the internal features between encrypted form and secret-shared form.

However, these hybrid secure inference solutions assume semi-honest participants. Such assumption is not guaranteed in real scenarios. We observe that by considering malicious behavior, the client can secretly shift the internal features during inference, and observe its effect on the final output of the model. With this additional opportunity of changing the intermediate data, in this paper we propose a general model extraction method called SEEK (Safe-Error Extraction attacK), with which the client can extract the model parameters, more effectively than the extraction attack on models without secure inference.

\section{Related Works}


A model extraction attack method attepmts to retrieve information about a remotely-deployed model, and consequently copy the model parameters, mimic the model's functionality, or infer information about its training data. For a classification model, the target inference service may return class labels, top-$k$ probabilities, logits (or equivalently, all class probabilities), or even some intermediate features and/or gradients, among which class labels contain minimal information, leading to the most secure setup.

Most existing model extraction attacks \cite{tramer2016stealing,papernot2017practical,jagielski2020high,carlini2020cryptanalytic} target traditional model inference service in plaintext, while \cite{lehmkuhl2021muse} and this work target encrypted model inference service. Extraction methods also differ in their objectives. 
We follow the taxonomy made in \cite{jagielski2020high}, which categorized the extraction objectives into the following types:
\begin{itemize}
\item Exact Extraction: extract all parameters of the target model. This objective is not possible for plaintext inference service, due the model's inherent symmetries. We will show it can be efficiently achieved for encrypted model inference service.
\item Functionally Equivalent Extraction: construct a model such that its output is identical with that of the target model. The extracted model has the same structure as the target model, and the same paramters up to a symmetry transformation. This is the highest possible objective against a plaintext inference service.
\item Fidelity Extraction: For some input data distribution $\mathcal{D}$ and some goal similarity function $S(\cdot,\cdot)$, Fidelity Extraction aims to construct a model $\hat O$, such that $\textrm{Pr}_{x\sim \mathcal{D}}[S(\hat O(x), O(x))]$ is maximized. Typically, Fidelity Extraction only guarantee the outputs from the constructed model and the target model are similar enough on some test dataset.
\item Task Accuracy Extraction: construct a model to match or exceed the accuracy of the target model.
\end{itemize}



Learning-based methods access the target model to generate a training dataset, with which a substitute model is trained. Typically, learning-based methods do not attempt to extract individual parameters, resulting in Fidelity Extraction as the objective, and is generally query-efficient. In \cite{papernot2017practical}, in order to find the decision boundaries between classes efficiently, an iterative training algorithm is proposed, which uses the substitute model to create samples close to decision boundaries. In \cite{jagielski2020high}, the authors leveraged several recent optimizations in training, including unlabelled training, distillation, rotation loss, and MixMatch, to be able to train a substitute model with much fewer queries than the size of the original training set.

Direct recovery methods, which aims Functionally Equivalent Extraction, treat the target model as a function explicitly expressed by the parameters, and attempt to solve for the parameters given model query inputs and outputs. In \cite{jagielski2020high}, for neural networks which use ReLU activation and return logits, an extraction algorithm is devised by solving the parameters in the model, which is a piecewise-linear function. In \cite{carlini2020cryptanalytic}, the authors utilized methodologies from cryptoanalysis, and carefully improved the differential extraction method in \cite{jagielski2020high}, by treating more efficiently the issues arising from larger depths and numerical errors. 

In \cite{lehmkuhl2021muse}, the target inference service is performed with the hybrid MPC-HE scheme, and is assumed to return logits. The extraction method shifts the features so that the inference becomes a linear system, whose paramters can be solved with enough query inputs and outputs. 


Features of the related works are summarized in table \ref{comparison}. In comparison, the proposed method SEEK considers the most restrictive setup in which only class labels are returned. Additionally, by utilizing the ``safe-error attack'' method \cite{yen2000checking,joye2002montgomery}, SEEK does not suffer from the numerical error induced by very deep networks, and can apply to models with arbitrary number of layers.

\begin{table}
\centering
\small
\begin{tabular}{ |c|c|c|c|c|c| }
\hline
Method & Extraction method & Extraction target & \specialcell{Model\\output} & \specialcell{Highest\\model depth} & \specialcell{\# model calls\\per parameter} \\ 
\hline
Tram{\`e}r \textsl{et al.} \cite{tramer2016stealing} & Learning & Functional Equivalence & logits & 2 & $0.5 \sim 5$ \\
Tram{\`e}r \textsl{et al.} \cite{tramer2016stealing} & Direct Recovery & Functional Equivalence & labels & 1 & $20 \sim 50$ \\
Tram{\`e}r \textsl{et al.} \cite{tramer2016stealing} & Learning & Fidelity & labels & 3 & $\sim 100$ \\
Papernot \textsl{et al.} \cite{papernot2017practical} & Learning & Fidelity & labels & unlimited & $<1$ \\
Jagielski \textsl{et al.} \cite{jagielski2020high} & Learning & Fidelity & labels & unlimited & $\ll 1$ \\
Jagielski \textsl{et al.} \cite{jagielski2020high} & Direct Recovery & Functional Equivalence & logits & 2 & $\sim 10$ \\ 
Carlini \textsl{et al.} \cite{carlini2020cryptanalytic} & Direct Recovery & Functional Equivalence & logits & 4 & $\sim 200$ \\ 
MUSE \cite{lehmkuhl2021muse} & MPC Malleation & Exact Extraction & logits & 10 & $1/n_c$ \\ 
SEEK & MPC Malleation & Exact Extraction & labels & unlimited & $\sim 50$ \\ 
\hline
\end{tabular}
\caption{Feature of the extraction methods against neural network and logistic regression models.}
\label{comparison}
\end{table}



It is well-known that the MPC protocol malleation attacks as in \cite{lehmkuhl2021muse} and this work can be mitigated by using a protocol with malicious security. Recently, a line of work with client-malicious model are proposed \cite{lehmkuhl2021muse, chandran2021simc, xu2022simc}. These protocols are designed based on authenticated shares, and are closing the gap of computational and communicational efficiency with respect to the protocols with semi-honest security.

\section{Secure inference setup}

We consider a general deep convolutional neural network (CNN), trained for a classification task. Layers in CNN can be categorized into linear layers and non-linear layers. Linear layers include convolution layers, fully-connected (FC) layers, as well as normalization layers, average-pooling layers. Addition and concatenation layers can be viewed as linear layers as well. Consecutive linear layers can be merged together to form a single linear layer. 
A linear layer indexed with $\ell$ in general can be expressed as
\begin{equation}
y_\ell = w_\ell\cdot x_\ell+b_\ell,
\label{linear_layer}
\end{equation}
where $x_\ell$ is the input feature map, $y_\ell$ is the output feature map, $w_\ell$ is the weight parameter, and $b_\ell$ is the bias parameter.
In this formalism, for a convolution layer, the weight parameters are sparse due to localized kernel, and the values are shared across spatial locations.

Non-linear layers include activation layers, which perform some element-wise nonlinear function, as well as max-pooling, softmax, and argmax layers. In this paper, for activation layers, we use the most common $\textrm{ReLU}$ activation. For our purpose, two adjacent non-linear layers will be viewed as a single non-linear layer. Typically for a CNN, a non-linear layer indexed with $\ell$ is composed of an activation layer
\begin{equation}
z_\ell = \textrm{ReLU}(y_\ell),
\label{relu_layer}
\end{equation}
or composed of a max-pooling layer followed by an activation layer,
\begin{equation}
z_\ell = \textrm{ReLU}(\textrm{maxpool}(y_\ell)),
\label{maxpool_relu_layer}
\end{equation}
or, for the last layer, composed of an argmax layer,
\begin{equation}
z_\ell=\textrm{argmax}(y_\ell),
\label{argmax}
\end{equation}
where $y_\ell$ is the input feature map, and $z_\ell$ is the output feature map.
We do not restrict the network structure to be linear, and structures such as skip connection and Inception are allowed. 

With hybrid secure inference solutions, the client encrypts its input $x_0$ into $\enc{x_0}$, and sends $\enc{x_0}$ to the server. For each linear layer as in equation (\ref{linear_layer}), the server computes
$$
\enc{y_\ell} = w_\ell\cdot \enc{x_\ell}+b_\ell.
$$
To perform a non-linear layer as in equation (\ref{relu_layer}), (\ref{maxpool_relu_layer}), and (\ref{argmax}), the server generates a random mask $r^y_\ell$, computes $\enc{y_\ell}-r^y_\ell=\enc{y_\ell-r^y_\ell}$, and send this encrypted value to the client. The client decrypts to get $y_\ell-r^y_\ell$. Now the two parties hold secret shares of the intermediate value $y_\ell$. The server and the client invoke a two-party MPC protocol corresponding to the non-linear layer. As the result, the client holds $z_\ell-r^z_\ell$, and the server holds $r^z_\ell$. To transform $z_\ell$ back to encrypted form, the client encrypts $z_\ell-r^z_\ell$ and sends $\enc{z_\ell-r^z_\ell}$ to the server, who can compute $\enc{z_\ell}$ and proceed to the next layer. After all layers are processed, the client and the server run another MPC protocol to compute equation (\ref{argmax}), and the client reconstructs the shares to get $c$.

Security of the hybrid protocol guarantees the privacy of input data, intermediate features, final result, as well as the model parameters, if the two parties follow the semi-honest model. However, we observe that the client can add arbitrary shifts to the secret shares in this protocol, and semantic security of the protocol ensures the server cannot detect the shift. Instead of using $y_\ell-r^y_\ell$ as input of the MPC calculation, the client can change it to $y_\ell-r^y_\ell+\delta y_\ell$, effectively changing the underlying value from $y_\ell$ to $y_\ell+\delta y_\ell$. Similarly, the client can change $z_\ell-r^z_\ell$ to $z_\ell-r^z_\ell+\delta z_\ell$, effectively changing the underlying value from $z_\ell$ to $z_\ell+\delta z_\ell$.

Thus the client is capable of shifting all inputs and outputs of the non-linear layers by arbitrary values, although the client is ignorant of the  values of the features. We consider how the client can exploit these additional inputs, to efficiently extract the model parameters. From the viewpoint of a malicious client, the model service can be formulated as
\begin{align*}
c&=C_{\{w\}}(x_0, \{\delta y_\ell,\delta z_\ell:\ell\in N\})\\
&=\textrm{argmax}(F_{\{w\}}(x_0, \{\delta y_\ell,\delta z_\ell:\ell\in N\})),
\end{align*}
where $\{w\}$ denotes all model parameters, $C$ is the functionality of the classification model, which outputs the predicted class index, $N$ is the set of non-linear layers, and $F$ is the output of the last linear layer.

To ease the notation, we use ``named arguments'' to denote the set of inputs as $(x_0, \{\delta y_\ell,\delta z_\ell:\ell\in N\}) = V(\widetilde{x_0}=x_0, \ldots, \widetilde{\delta y_\ell}=\delta y_\ell,\ldots, \widetilde{\delta z_{\ell'}}=\delta z_{\ell'},\ldots)$. If an input is not present in the list of arguments of $v$, it means the input is set to zeros. For example, $V(\widetilde{\delta y_\ell}=\delta y_\ell) = V(\widetilde{x_0}=0, \widetilde{\delta y_\ell}=\delta y_\ell, \widetilde{\delta y_{\ell'}}=0, \widetilde{\delta z_{\ell''}}=0)$, for all $\ell'\neq\ell$ and all $\ell''$. Two sets of inputs can be added with the natural element-wise addition.

Because the output of the model is a discrete value, in order to extract model parameters, the adversary needs to find the boundary between classes, where for
$$
y_{\ell_\textrm{last}} = F_{\{w\}}(v),
$$
it satisfies
\begin{equation}
y_{\ell_\textrm{last},c_1}=y_{\ell_\textrm{last},c_2}
\label{critical_condition_equal}
\end{equation}
for two different classes $c_1, c_2$, and 
\begin{equation}
y_{\ell_\textrm{last},c_1} > y_{\ell_\textrm{last}, c'}
\label{critical_condition_greater}
\end{equation}
for all other $c'$. In the following, we call a set of input satisfying the above relations a critical point, and denote the corresponding input variables with a $*$ subscript.

Starting from a set of inputs and changing feature values on a layer, a critical point can always be found. Algorithm \ref{bisection} shows a routine for finding a critical point using bisection, in which all input variables are fixed except $\delta y_\ell$. 

\section{Extraction of intermediate features}

In this section, we present the concrete method of SEEK. The adversary is able to shift all the inputs and outputs of the activation layers, and observe the effect on the model output. One way to extract the parameters is to find the space of critical points formed by shifting the intermediate features. However, because the landscape of model output as a function of the shifts can be very complicated, this method becomes intractable when the target layer is far away from the output layer. Instead, starting from a critical point, the proposed method will add a particular set of shifts, such that if the corresponding feature satisfies certain condition, the added shifts would cancel itself and do not affect any other features. We can test the criticality of the shifted input, and determine the value of the target feature. In this way, we keep the effect of the shifts to a minimal level, making this extraction method numerically stable and the errors in the extracted parameters independent of each other. This extraction strategy is in concept similar with the safe-error attack \cite{yen2000checking,joye2002montgomery} as a type of fault injection attack to security systems.

\begin{algorithm}
\caption{$\textsf{search\_critical}$ -- Find a critical point by shifting $y_\ell$ from a given set of inputs.}
\label{bisection}
\SetKwInOut{KwIn}{Input}
\SetKwInOut{KwOut}{Output}
\SetKwRepeat{Do}{do}{while}
\KwIn{A fixed set of inputs $v^0$, variable input layer index $\ell$, norm $d$, and error threshold $\epsilon$}
\KwOut{$\delta y^*_\ell$}
\Do{$c^1=c^2$}{
Randomly sample $\delta y^1_\ell$ and $\delta y^2_\ell$ with norm $d$\;
$c^1\leftarrow C\left(v^0 + V(\widetilde{\delta y_\ell}=\delta y^1_\ell)\right)$, $c^2\leftarrow C\left(v^0 + V(\widetilde{\delta y_\ell}=\delta y^2_\ell)\right)$\;
}
\While{$|\delta y^2_\ell-\delta y^1_\ell| > \epsilon$}{
  $\delta y^3_\ell \leftarrow (\delta y^1_\ell + \delta y^2_\ell) / 2$, and normalize $\delta y^3_\ell$ with norm $d$\;
  $c^3\leftarrow C\left(v^0 + V(\widetilde{\delta y_\ell}=\delta y^3_\ell)\right)$\;
  \eIf{$c^3=c^1$}{
  $\delta y^1_\ell\leftarrow\delta y^3_\ell$\;
  }{
  $\delta y^2_\ell\leftarrow\delta y^3_\ell$, $c^2\leftarrow c^3$\;
  }
}
\Return{$\delta y^*_\ell \leftarrow \delta y^2_\ell$\;}
\end{algorithm}

\subsection{Extraction of standalone ReLU layer inputs}

In this subsection, we present the method to extract an input feature value of a standalone ReLU activation as in equation (\ref{relu_layer}).

Consider a critical point $v^*$. A target feature $y_{\ell, i}$, which is the input of a standalone ReLU activation, takes the value $y^*_{\ell, i}$ from the set of input $v^*$. If $y^*_{\ell, i}<0$, shifting it by a small positive or any negative $\delta y_{\ell, i}$ will not affect the model output, because $\textrm{ReLU}(y^*_{\ell, i}+\delta y_{\ell, i}) =  \textrm{ReLU}(y^*_{\ell, i}) = 0$. In this case we add a positive shift to $y_{\ell, i}$. If $\delta y_{\ell, i}$ is large enough such that $y^*_{\ell, i}+\delta y_{\ell, i}>0$, which implies $\textrm{ReLU}(y^*_{\ell, i}+\delta y_{\ell, i})\neq \textrm{ReLU}(y^*_{\ell, i})$, the input is no longer a critical point.  We can test the criticality of the input while varying $\delta y_{\ell, i}$. At the boundary between critical points and non-critical points, $y^*_{\ell,i} = -\delta y_{\ell,i}$.


On the other hand, if $y^*_{\ell,i}>0$, subtracting a small positive or any negative $\delta y_{\ell, i}$ from $y_{\ell, i}$, and at the same time adding the same shift $\delta y_{\ell, i}$ to $z_{\ell, i}$, will not affect the model output, because $\textrm{ReLU}(y^*_{\ell, i}-\delta y_{\ell, i})+\delta y_{\ell, i} = \textrm{ReLU}(y^*_{\ell, i}) = y^*_{\ell, i}$. If $\delta y_{\ell, i}$ is large enough such that $y^*_{\ell, i}-\delta y_{\ell, i}<0$, the input is no longer a critical point. We can test the criticality of the input while varying $\delta y_{\ell, i}$. At the boundary between critical points and non-critical points, $y^*_{\ell,i} = \delta y_{\ell,i}$.

To test the criticality of a set of inputs $v$, we use the properties equation (\ref{critical_condition_equal}) and (\ref{critical_condition_greater}). Consider we start from a critical point $v^*$ at the boundary between class $c_1$ and $c_2$, and add some shifts $\delta v$ to $y_\ell^*$ and $z_\ell^*$. If the added shifts do not change any feature values other than $y_\ell$ and $z_\ell$, then the set of inputs $v=v^*+\delta v$ is also a critical point between class $c_1$ and $c_2$. In this case, $v$ satisfy
$$
C\left(v + V(\widetilde{\delta y_{\textrm{last}, c_1}} = \epsilon)\right) = c_1,
$$
and
$$
C\left(v + V(\widetilde{\delta y_{\textrm{last}, c_2}} = \epsilon)\right) = c_2,
$$
where $\epsilon$ is a small positive value. If the added shifts affect other feature values, then the above equations are not satisfied with overwhelming probability.


The algorithm for extracting a input feature of a standalone ReLU is shown in algorithm \ref{extract_feature}.

\begin{algorithm}
\caption{$\textsf{extract\_feature}$ -- Extraction of an intermediate feature value at a critical point}
\label{extract_feature}
\SetKwInOut{KwIn}{Input}
\SetKwInOut{KwOut}{Output}
\SetKwRepeat{Do}{do}{while}
\KwIn{Input critical point $v^*$, target activation layer index $\ell$, and target feature index $i$}
\KwOut{$y^*_{\ell,i}$}
\eIf{$v = v^* + V(\widetilde{\delta y_{\ell,i}}=-1)$ is critical}{
\tcp{$y^*_{\ell, i}\le 0$}
For points of the form $v = v^* + V(\widetilde{\delta y_{\ell,i}}=\eta)$, where $\eta\in[0,\infty)$, search for the boundary $\eta=\bar\eta$ between critical points and non-critical points\;
\Return{$y^*_{\ell,i} \leftarrow -\bar\eta$}\;
}{
\tcp{$y^*_{\ell, i}>0$}
For points of the form $v = v^* + V(\widetilde{\delta y_{\ell,i}}=-\eta, \widetilde{\delta z_{\ell,i}}=\eta)$, where $\eta\in[0,\infty)$, search for the boundary $\eta=\bar\eta$ between critical points and non-critical points\;
\Return{$y^*_{\ell,i} \leftarrow \bar\eta$}\;
}
\end{algorithm}

\subsection{Extraction of maxpool-ReLU layer inputs}

In this subsection, we present the method to extract an input feature value of a maxpool layer followed by a ReLU layer, as in equation (\ref{maxpool_relu_layer}).

The method is similar with the one in previous subsection. Because the maxpool layer maps multiple features into one feature, when adjusting one input feature and one output feature, we need to find a way to suppress the effect of the other related input features.

\begin{figure}
\centering
\includegraphics[scale=0.30]{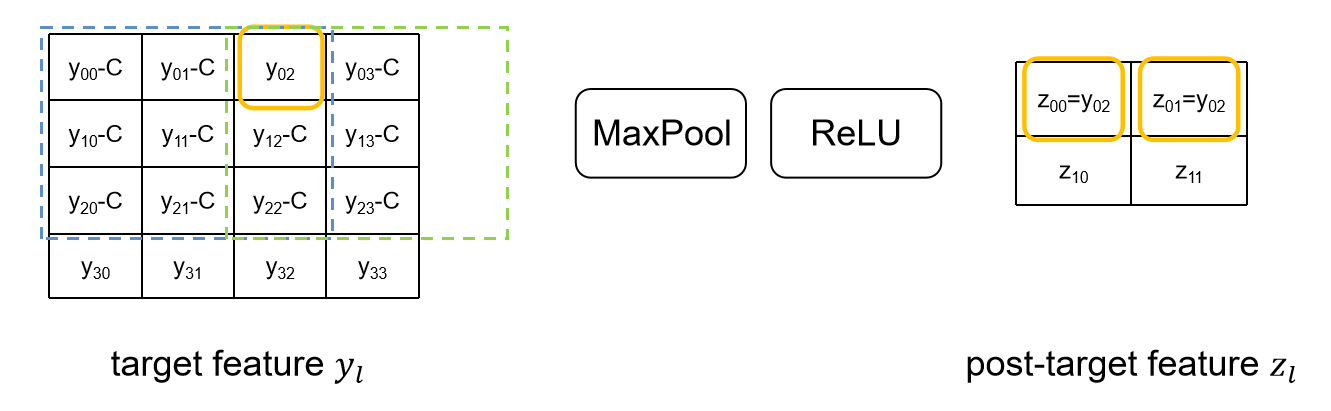}
\caption{An example of extraction method for a maxpool-ReLU layer input feature. Features in the orange boxes are the target feature $y_{\ell,i}$ and its related features $Z_{\ell,i}$ in the post-target layer, respectively. The dashed rectangles are ranges of maxpool kernels. $C$ is a large positive constant, added in order to suppress the effect from other features on $y_\ell$ to $Z_{\ell,i}$.}
\label{maxpool_relu}
\end{figure}

Assume the feature value to be extracted is $y_{\ell,i}$, and the set of output features affected by shifting $y_{\ell,i}$ is $Z_{\ell,i}$. We can add a large negative shift to all features in $y_\ell$, except $y_{\ell,i}$. As a result, $z_\ell$ will be zero everywhere except features in $Z_{\ell,i}$, which takes the value of $y_{\ell,i}$. Now we can shift the value of $y_{\ell,i}$ and values in $Z_{\ell,i}$, observe the effect on the criticality, and consequently extract $y_{\ell,i}$. The extraction process is similar with algorithm \ref{extract_feature}, except now the feature $z_{\ell,i}$ is replaced by a set of features which should be shifted together. See figure \ref{maxpool_relu} for an illustration of this method.

\subsection{Extraction of linear layer parameters}

The methods presented in the previous two subsections can extract all the intermediate features of a critical point. Then for each linear layer as in equation (\ref{linear_layer}), with the input features and output features known, the formula is a set of linear equations for $w_\ell$ and $b_\ell$. We can repeat this process and collect enough equations to solve all the model parameters.

To further simplify the extraction process, we note that we can add a large negative shift to the input of a ReLU activation, and ensure its output to be zero. We can also add arbitrary shifts to the zeroed outputs. Thus we have a means to accurately control the output values of ReLU activations. In equation (\ref{linear_layer}), by setting $x_\ell$ to be identically zero and extracting $y_{\ell}$, the value of $b_\ell$ can be read off,
$$
b_{\ell,j} = y_{\ell,j},
$$
where $j$ is an output feature index. By setting all but one feature value of $x_\ell$ zero and extracting $y_{\ell}$, the weight parameters can be derived as,
$$
w_{\ell,j,i_0} = \frac{y_{\ell,j} - b_{\ell,j}}{x_{\ell, i_0}},
$$
where $i_0$ is index of non-zero $x_\ell$ value.

We observe that algorithm \ref{extract_feature} can work on multiple target feature indices, if all the target features at these indices have the same value. In practice, running algorithm \ref{extract_feature} on more indices improves the accuracy, because the influence of a change of their value is more significant to the model output. For convolutional layers, we can use its structure to create multiple target features with the same value. For the bias, by setting $x_\ell$ to be identically zero, all values on $y_{\ell, c_{\textrm{out}}}$ are equal to $b_{\ell,c_{\textrm{out}}}$, where $c_{\textrm{out}}$ is an output channle index. For the weight, instead of setting one feature value on $x_\ell$ nonzero, for an input channel index $c_{\textrm{in}}$, we can set $x_{\ell,c_{\textrm{in}}}$ to be periodically nonzero, so that the target kernel value is repeated on $y_{\ell}$.

\begin{figure}
\centering
\includegraphics[scale=0.30]{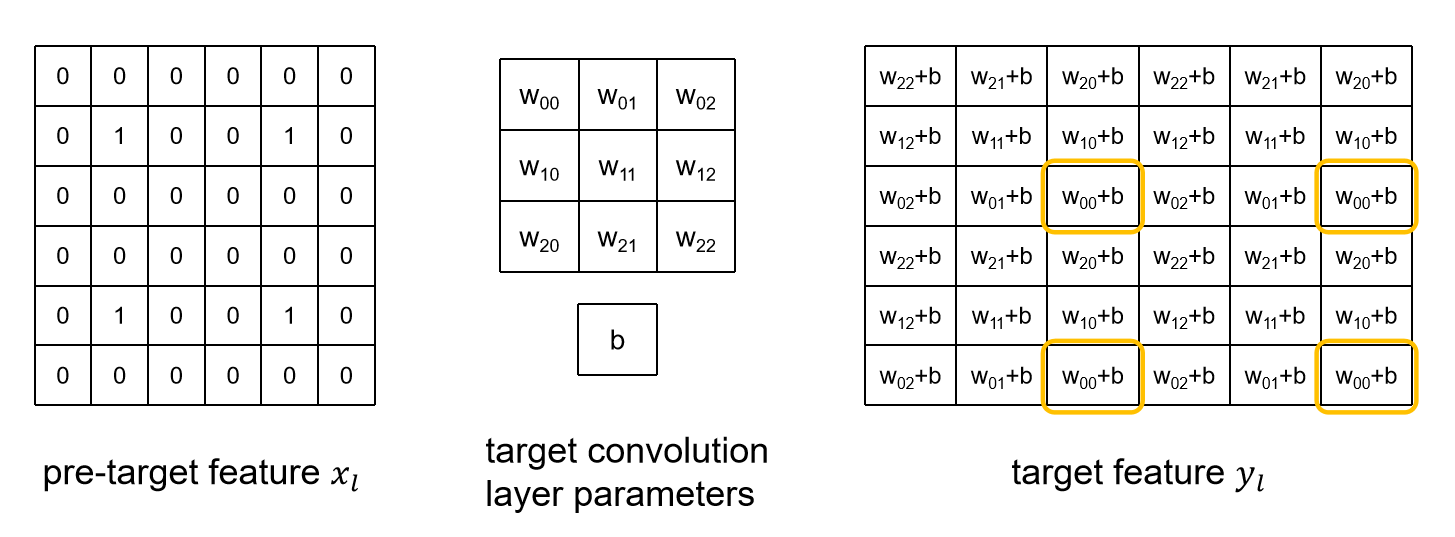}
\caption{An example of the convolutin layer extraction method, as shown in algorithm \ref{full_extraction}. For simplicity, only one channel for each layer is shown. By setting the pre-target feature $x_\ell$ to be nonzero with a period of kernel size, the target feature layer $y_\ell$ is also periodic, and the values in the orange boxes can be extracted together for better accuracy, which reveal the values of target convolution layer parameters.}
\label{example}
\end{figure}

The above extraction process is illustrated in figure \ref{example}. Algorithm \ref{full_extraction} shows the complete algorithm to extract the parameters in a convolution layer. For clarity, we assume that the stride of the convolution to be 1. The extraction algorithm for a fully-connected layer is similar, and is omitted for brevity.

\subsection{Extraction of last linear layer parameters}

The extraction method described in the previous subsection applies to all the linear layers, except the last fully-connected layer before the argmax layer. Without a ReLU layer after the last fully-connected layer, the features $y_{\textrm{last}}$ cannot be extracted with the $\textsf{extract\_feature}$ routine. Instead, the following method can be applied. Assume the numbers of input and output features of the last fully-connected layer are $n_0$ and $n_1$, respectively. To extract $b_{\textrm{last}}$, we can add shifts to the layer before the last layer, so that $x_{\textrm{last}}=0$. Then we search for critical points by varying $\delta y_{\textrm{last}}$, which gives the relation about $b_{\textrm{last}}$,
$$
b_{\textrm{last},c_1} + \delta y^*_{\textrm{last},c_1} = b_{\textrm{last},c_2} + \delta y^*_{\textrm{last},c_2}.
$$
$n_1-1$ such equations give the values of $b_{\textrm{last}}$ up to an additive constant. Similarly for $w_{\textrm{last}}$, we can manipulate the layer before the last layer, so that $x_{\textrm{last}}$ is zero except at feature $i_0$. Then we search for critical points by varying $\delta y_{\textrm{last}}$, which gives the relation about $w_{\textrm{last}}$,
\begin{align*}
w_{\textrm{last},c_1,i_0}x_{\textrm{last},i_0} + b_{\textrm{last},c_1} + \delta y^*_{\textrm{last},c_1} &= w_{\textrm{last},c_2,i_0}x_{\textrm{last},i_0} + b_{\textrm{last},c_2} + \delta y^*_{\textrm{last},c_2},\\
w_{\textrm{last},c_1,i_0} - w_{\textrm{last},c_2,i_0} &= \frac{(b_{\textrm{last},c_2} - b_{\textrm{last},c_1}) + \delta y^*_{\textrm{last},c_2} - \delta y^*_{\textrm{last},c_1}}{x_{\textrm{last},i_0}}.
\end{align*}
$(n_1-1)n_0$ such equations gives the values of $w_{\textrm{last}}$, up to $n_0$ additive constants. In fact, because only the class label is observed, this is all the degrees of freedom of $w_{\textrm{last}}$ that can be determined.

\begin{algorithm}
\caption{Extraction of parameters in a convolution layer}
\label{full_extraction}
\SetKwInOut{KwIn}{Input}
\SetKwInOut{KwOut}{Output}
\SetKwRepeat{Do}{do}{while}
\KwIn{Target convolution layer index $\ell$, numbers of output and input channels $n_{\textrm{out}}$ and $n_{\textrm{in}}$, convolution kernal size $(k_h, k_w)$, input feature size $(f_h, f_w)$}
\KwOut{Target convolution layer parameters $b_{\ell}$ and $w_{\ell}$}
Get the layer index $\ell_0$ whose output is the input of layer $\ell$, i.e., $z_{\ell_0}=x_\ell$\;
Get the index of last non-linear layer $\ell_{\textrm{last}}$\;
Add a large negative shift $-d$ to all features in $y_{\ell_0}$\;
$\delta z^*_{\ell_{\textrm{last}}} \leftarrow \textsf{search\_critical}(V(\widetilde{x_0}=x_0, \widetilde{\delta y_{\ell_0}}=-d), \ell_{\textrm{last}})$\;
$v^* \leftarrow V(\widetilde{x_0}=x_0, \widetilde{\delta y_{\ell_0}}=-d, \widetilde{\delta z_{\ell_{\textrm{last}}}}=\delta z^*_{\ell_{\textrm{last}}})$\;
\For{$c_{\textrm{out}}\leftarrow 0$ \KwTo $n_{\textrm{out}}-1$}{
$\beta \leftarrow \{(c_{\textrm{out}}, i, j): 0\le i< f_h, 0\le j < f_w\}$\;
$b_{\ell,c_{\textrm{out}}}\leftarrow \textsf{extract\_feature}(v^*, \ell, \beta)$\;
}
$k_h'\leftarrow (k_h-1)/2, k_w'\leftarrow (k_w-1)/2$\;
$\Delta \leftarrow (n_\textrm{in} \cdot k_h \cdot k_w / 4)^{1/2}$\;
\For{$c_{\textrm{in}}\leftarrow 0$ \KwTo $n_{\textrm{in}}-1$}{
Create a feature map $\alpha$ of size $(f_h, f_w)$ whose values are $\alpha_{i, j} = \Delta \cdot \hat\delta((i-k_h')\%k_h) \cdot \hat\delta((j-k_w')\%k_w))$, where $\hat\delta(\cdot)$ is the discrete delta function\;
$\delta z^*_{\ell_{\textrm{last}}} \leftarrow \textsf{search\_critical}(V(\widetilde{x_0}=x_0, \widetilde{\delta y_{\ell_0}}=-d, \widetilde{\delta x_{\ell, c_{\textrm{in}}}} = \alpha, \ell_{\textrm{last}})$\;
$v^* \leftarrow V(\widetilde{x_0}=x_0, \widetilde{\delta y_{\ell_0}}=-d, \widetilde{\delta x_{\ell, c_{\textrm{in}}}} = \alpha, \widetilde{\delta z_{\ell_{\textrm{last}}}}=\delta z^*_{\ell_{\textrm{last}}})$\;
\For{$c_{\textrm{out}}\leftarrow 0$ \KwTo $n_{\textrm{out}}-1$}{
\For{$i\leftarrow 0$ \KwTo $k_h-1$}{
\For{$j\leftarrow 0$ \KwTo $k_w-1$}{
$\beta\leftarrow\{(c_{\textrm{out}}, i',j'):0\le i'<f_h,(i'-k_h+1+i)\%k_h=0,0\le j'<f_w,(j'-k_w+1+j)\%k_w=0\}$\;
$y_{\ell, c_{\textrm{out}}, i, j}\leftarrow \textsf{extract\_feature}(v^*, \ell, \beta)$\;
$w_{\ell, c_{\textrm{out}}, c_{\textrm{in}}, i, j} \leftarrow (y_{\ell, c_{\textrm{out}}, i, j} - b_{\ell,c_{\textrm{out}}}) / \Delta$\;
}
}
}
}
\Return{$b_{\ell}$ and $w_{\ell}$}\;
\end{algorithm}

\section{Experiment}

We test the proposed SEEK method on ResNet-18 \cite{he2016deep}, implemented in the latest PyTorch \cite{NEURIPS2019_9015} release. The model contains 11.7M parameters, and is trained for ImageNet classification task.

\begin{figure}
\centering
\includegraphics[scale=0.35, angle=0]{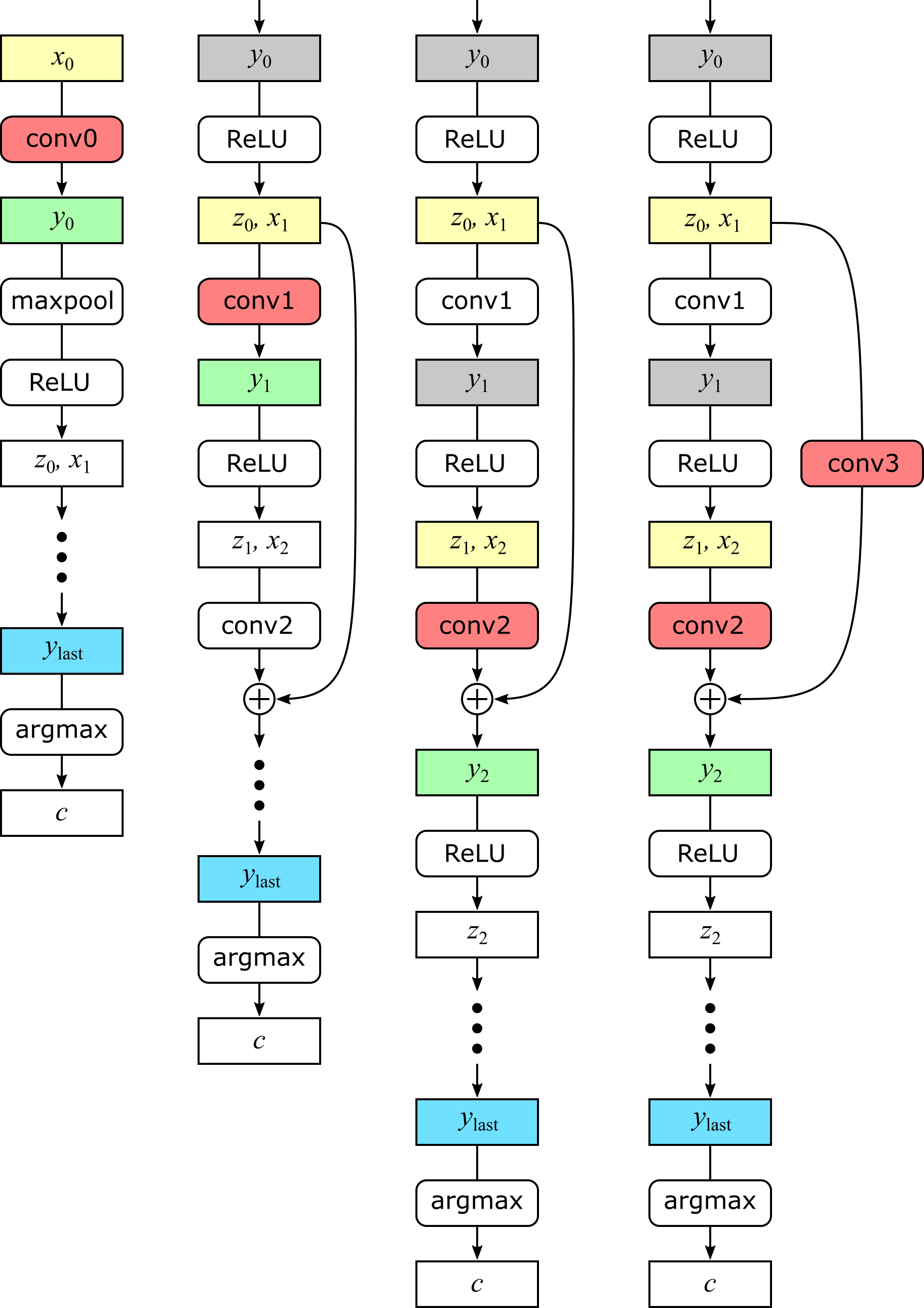}
\caption{Examples of extraction paths for ResNet. A large negative value is added to the grey layers, so that the feature values in the pre-target layers (yellow) can be adjusted to some convenient values. A bisection search is performed on the last feature layers (blue) to find a critical point. The feature values of the post-target layers (green) are extracted, based on properties of the non-linear succeeding non-linear layers. Then the parameters of the target layers (red) are extracted. }
\label{extraction_paths}
\end{figure}

In ResNet-18, some of the linear layers have a single preceding layer, while the layers immediately after the addition layers have two preceding layers. In addition, some skip connections are identity connections, and some are down-sampling connections, which have their own convolution weights. In all of these cases, we can use the methods in the previous section to extract the linear layers' parameters. Figure \ref{extraction_paths} shows several extraction paths for different cases in ResNet.

\begin{figure}
\centering
\includegraphics[scale=0.7, angle=0]{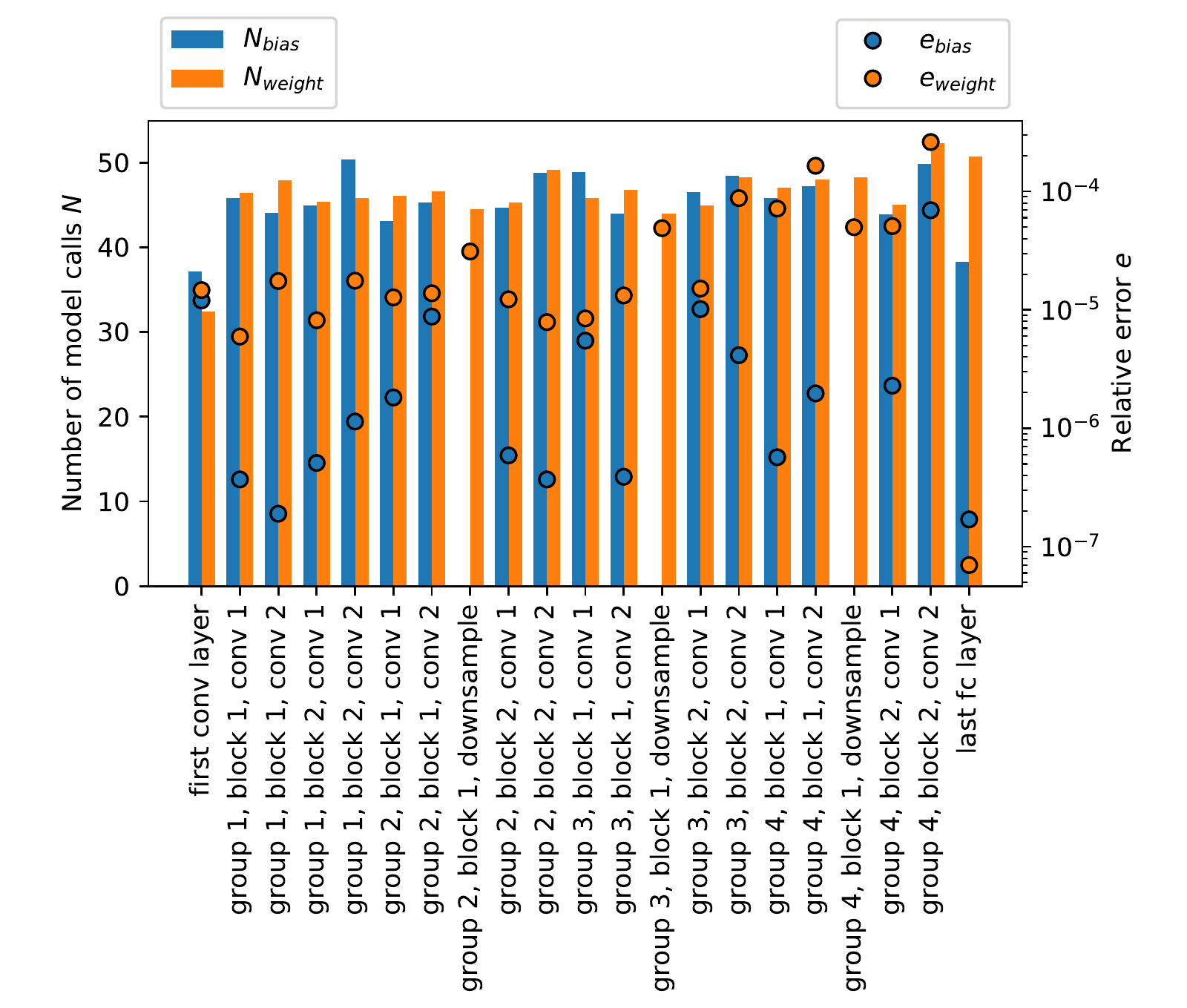}
\caption{Result of the proposed extraction method on ResNet-18. $N_{\textrm{bias}}$($N_{\textrm{weight}}$) is the average number of model calls for extracting a bias(weight) parameter. $e_{\textrm{bias}}$($e_{\textrm{weight}}$) is the average relative error of the extracted bias(weight) parameter.}
\label{experiment}
\end{figure}

We implemented the extraction algorithm, and experimentally tested its performance. Figure \ref{experiment} shows the average number of model calls required for extracting each parameter, as well as the average relative error, for different layers in ResNet-18. For each parameter, the average number of model calls is $45.8$. The average relative error of bias is $6.68\times 10^{-6}$, and the average relative error of weight is $4.35\times 10^{-5}$.

As figure \ref{experiment} shows, the average error of weight tends to be larger as the layer is closer to the model output, except for the last FC layer. The reason for this phenomenon is, if the shift of the target feature is larger than its value (see algorithm \ref{extract_feature}), the output of ReLU function $z=z^*+\delta z$ will be different from its original value $z^*$ and in turn changes the final logits. However, for a layer closer to the model output, the relationship between $\delta z$ and the final logits $y_{\textrm{last},c_1}$ and $y_{\textrm{last},c_2}$ becomes simpler. In some rare cases, $\delta z$ affects $y_{\textrm{last},c_1}$ and $y_{\textrm{last},c_2}$ approximately in the same way in a small neighborhood. In this small neighborhood of $\delta z$ value, algorithm \ref{extract_feature} cannot distinct the shift by criticality test, and resulting in a larger error. This issue can be mitigated by repeating the extraction multiple times with different initial critical points.

\section{Conclusion and discussion}

In this work, we proposed SEEK, a model extraction attack method against HE-MPC hybrid inference service with semi-honest security, with the most stringent assumption that the model outputs class labels only. Our method makes use of the piecewise-linear property of the ReLU activation, and the principle of safe-error attack, thus achieving an extraction process that can accurately extract each layer's parameters. As the method tests whether a shift to the internal feature affects the criticality of the whole input, it is not affected by the depth of the model, which can incur numerical issues for other extraction methods. Furthermore, because the extraction of parameters in a layer is not dependent on the extraction result of any other layer, a distributed extraction attack is straightforward.

SEEK can be generalized to other secure inference protocols with semi-honest security. In particular, if the ReLU activation function is replaced by other piecewise-linear functions, such as ReLU6 or leaky ReLU, our method can be applied in essentially the same manner. If the activation function is linear only in part of the input range, such as the swish activation, we can also manipulate the input so that it falls in the region of linear activation. For secure inference of decision tree models, the general method of safe-error attack is applicable, because the discrete nature of decision tree inference makes it possible to change individual intemediate feature and observe the effect on the final output. We leave the  security analysis of the case of decision tree models for future work.

As demonstrated by the proposed extraction method, the capability of changing all the intermediate features with arbitrary shifts is quite powerful, and it is non-trivial to prevent such attack. Shuffling the features in a layer before the MPC protocol only increases the difficulty of this attack by a constant factor. The model inference protocols with client-malicious security \cite{lehmkuhl2021muse, chandran2021simc, xu2022simc}, albeit with significant communicational and computational cost, provide systematic countermeasure against our attack. Secure inference based on fully homomorphic encryption with bootstrapping \cite{chillotti2021programmable, lee2022privacy}, or garbled circuits \cite{rouhani2018deepsecure,ball2019garbled,riazi2019xonn}, lead to another direction of mitigation, in which the inference is processed in constant communication rounds, so an adversary do not have the oppotunity to malleate intermediate model features.

\bibliographystyle{IEEEtran}

\begin{thebibliography}{10}
\providecommand{\url}[1]{#1}
\csname url@samestyle\endcsname
\providecommand{\newblock}{\relax}
\providecommand{\bibinfo}[2]{#2}
\providecommand{\BIBentrySTDinterwordspacing}{\spaceskip=0pt\relax}
\providecommand{\BIBentryALTinterwordstretchfactor}{4}
\providecommand{\BIBentryALTinterwordspacing}{\spaceskip=\fontdimen2\font plus
\BIBentryALTinterwordstretchfactor\fontdimen3\font minus
  \fontdimen4\font\relax}
\providecommand{\BIBforeignlanguage}[2]{{%
\expandafter\ifx\csname l@#1\endcsname\relax
\typeout{** WARNING: IEEEtran.bst: No hyphenation pattern has been}%
\typeout{** loaded for the language `#1'. Using the pattern for}%
\typeout{** the default language instead.}%
\else
\language=\csname l@#1\endcsname
\fi
#2}}
\providecommand{\BIBdecl}{\relax}
\BIBdecl

\bibitem{fredrikson2015model}
M.~Fredrikson, S.~Jha, and T.~Ristenpart, ``Model inversion attacks that
  exploit confidence information and basic countermeasures,'' in
  \emph{Proceedings of the 22nd ACM SIGSAC conference on computer and
  communications security}, 2015, pp. 1322--1333.

\bibitem{shokri2017membership}
R.~Shokri, M.~Stronati, C.~Song, and V.~Shmatikov, ``Membership inference
  attacks against machine learning models,'' in \emph{2017 IEEE symposium on
  security and privacy (SP)}.\hskip 1em plus 0.5em minus 0.4em\relax IEEE,
  2017, pp. 3--18.

\bibitem{lowd2005adversarial}
D.~Lowd and C.~Meek, ``Adversarial learning,'' in \emph{Proceedings of the
  eleventh ACM SIGKDD international conference on Knowledge discovery in data
  mining}, 2005, pp. 641--647.

\bibitem{gilad2016cryptonets}
R.~Gilad-Bachrach, N.~Dowlin, K.~Laine, K.~Lauter, M.~Naehrig, and J.~Wernsing,
  ``Cryptonets: Applying neural networks to encrypted data with high throughput
  and accuracy,'' in \emph{International conference on machine learning}.\hskip
  1em plus 0.5em minus 0.4em\relax PMLR, 2016, pp. 201--210.

\bibitem{brutzkus2019low}
A.~Brutzkus, R.~Gilad-Bachrach, and O.~Elisha, ``Low latency privacy preserving
  inference,'' in \emph{International Conference on Machine Learning}.\hskip
  1em plus 0.5em minus 0.4em\relax PMLR, 2019, pp. 812--821.

\bibitem{rouhani2018deepsecure}
B.~D. Rouhani, M.~S. Riazi, and F.~Koushanfar, ``Deepsecure: Scalable
  provably-secure deep learning,'' in \emph{Proceedings of the 55th annual
  design automation conference}, 2018, pp. 1--6.

\bibitem{ball2019garbled}
M.~Ball, B.~Carmer, T.~Malkin, M.~Rosulek, and N.~Schimanski, ``Garbled neural
  networks are practical,'' \emph{Cryptology ePrint Archive}, 2019.

\bibitem{riazi2019xonn}
M.~S. Riazi, M.~Samragh, H.~Chen, K.~Laine, K.~Lauter, and F.~Koushanfar,
  ``$\{$XONN$\}$:$\{$XNOR-based$\}$ oblivious deep neural network inference,''
  in \emph{28th USENIX Security Symposium (USENIX Security 19)}, 2019, pp.
  1501--1518.

\bibitem{mohassel2018aby3}
P.~Mohassel and P.~Rindal, ``Aby3: A mixed protocol framework for machine
  learning,'' in \emph{Proceedings of the 2018 ACM SIGSAC conference on
  computer and communications security}, 2018, pp. 35--52.

\bibitem{wagh2019securenn}
S.~Wagh, D.~Gupta, and N.~Chandran, ``Securenn: 3-party secure computation for
  neural network training.'' \emph{Proc. Priv. Enhancing Technol.}, vol. 2019,
  no.~3, pp. 26--49, 2019.

\bibitem{kumar2020cryptflow}
N.~Kumar, M.~Rathee, N.~Chandran, D.~Gupta, A.~Rastogi, and R.~Sharma,
  ``Cryptflow: Secure tensorflow inference,'' in \emph{2020 IEEE Symposium on
  Security and Privacy (SP)}.\hskip 1em plus 0.5em minus 0.4em\relax IEEE,
  2020, pp. 336--353.

\bibitem{juvekar2018gazelle}
C.~Juvekar, V.~Vaikuntanathan, and A.~Chandrakasan, ``$\{$GAZELLE$\}$: A low
  latency framework for secure neural network inference,'' in \emph{27th USENIX
  Security Symposium (USENIX Security 18)}, 2018, pp. 1651--1669.

\bibitem{boemer2019ngraph}
F.~Boemer, A.~Costache, R.~Cammarota, and C.~Wierzynski, ``ngraph-he2: A
  high-throughput framework for neural network inference on encrypted data,''
  in \emph{Proceedings of the 7th ACM Workshop on Encrypted Computing \&
  Applied Homomorphic Cryptography}, 2019, pp. 45--56.

\bibitem{mishra2020delphi}
P.~Mishra, R.~Lehmkuhl, A.~Srinivasan, W.~Zheng, and R.~A. Popa, ``Delphi: A
  cryptographic inference service for neural networks,'' in \emph{29th USENIX
  Security Symposium (USENIX Security 20)}, 2020, pp. 2505--2522.

\bibitem{rathee2020cryptflow2}
D.~Rathee, M.~Rathee, N.~Kumar, N.~Chandran, D.~Gupta, A.~Rastogi, and
  R.~Sharma, ``Cryptflow2: Practical 2-party secure inference,'' in
  \emph{Proceedings of the 2020 ACM SIGSAC Conference on Computer and
  Communications Security}, 2020, pp. 325--342.

\bibitem{huang2022cheetah}
Z.~Huang, W.-j. Lu, C.~Hong, and J.~Ding, ``Cheetah: Lean and fast secure
  two-party deep neural network inference.'' \emph{IACR Cryptol. ePrint Arch.},
  vol. 2022, p. 207, 2022.

\bibitem{tramer2016stealing}
F.~Tram{\`e}r, F.~Zhang, A.~Juels, M.~K. Reiter, and T.~Ristenpart, ``Stealing
  machine learning models via prediction $\{$APIs$\}$,'' in \emph{25th USENIX
  security symposium (USENIX Security 16)}, 2016, pp. 601--618.

\bibitem{papernot2017practical}
N.~Papernot, P.~McDaniel, I.~Goodfellow, S.~Jha, Z.~B. Celik, and A.~Swami,
  ``Practical black-box attacks against machine learning,'' in
  \emph{Proceedings of the 2017 ACM on Asia conference on computer and
  communications security}, 2017, pp. 506--519.

\bibitem{jagielski2020high}
M.~Jagielski, N.~Carlini, D.~Berthelot, A.~Kurakin, and N.~Papernot, ``High
  accuracy and high fidelity extraction of neural networks,'' in \emph{29th
  USENIX Security Symposium (USENIX Security 20)}, 2020, pp. 1345--1362.

\bibitem{carlini2020cryptanalytic}
N.~Carlini, M.~Jagielski, and I.~Mironov, ``Cryptanalytic extraction of neural
  network models,'' in \emph{Annual International Cryptology Conference}.\hskip
  1em plus 0.5em minus 0.4em\relax Springer, 2020, pp. 189--218.

\bibitem{lehmkuhl2021muse}
R.~Lehmkuhl, P.~Mishra, A.~Srinivasan, and R.~A. Popa, ``Muse: Secure inference
  resilient to malicious clients,'' in \emph{30th USENIX Security Symposium
  (USENIX Security 21)}, 2021, pp. 2201--2218.

\bibitem{yen2000checking}
S.-M. Yen and M.~Joye, ``Checking before output may not be enough against
  fault-based cryptanalysis,'' \emph{IEEE Transactions on computers}, vol.~49,
  no.~9, pp. 967--970, 2000.

\bibitem{joye2002montgomery}
M.~Joye and S.-M. Yen, ``The montgomery powering ladder,'' in
  \emph{International workshop on cryptographic hardware and embedded
  systems}.\hskip 1em plus 0.5em minus 0.4em\relax Springer, 2002, pp.
  291--302.

\bibitem{chandran2021simc}
N.~Chandran, D.~Gupta, S.~L.~B. Obbattu, and A.~Shah, ``Simc: Ml inference
  secure against malicious clients at semi-honest cost,'' \emph{Cryptology
  ePrint Archive}, 2021.

\bibitem{xu2022simc}
G.~Xu, X.~Han, T.~Zhang, H.~Li, and R.~H. Deng, ``Simc 2.0: Improved secure ml
  inference against malicious clients,'' \emph{arXiv preprint
  arXiv:2207.04637}, 2022.

\bibitem{he2016deep}
K.~He, X.~Zhang, S.~Ren, and J.~Sun, ``Deep residual learning for image
  recognition,'' in \emph{Proceedings of the IEEE conference on computer vision
  and pattern recognition}, 2016, pp. 770--778.

\bibitem{NEURIPS2019_9015}
\BIBentryALTinterwordspacing
A.~Paszke, S.~Gross, F.~Massa, A.~Lerer, J.~Bradbury, G.~Chanan, T.~Killeen,
  Z.~Lin, N.~Gimelshein, L.~Antiga, A.~Desmaison, A.~Kopf, E.~Yang, Z.~DeVito,
  M.~Raison, A.~Tejani, S.~Chilamkurthy, B.~Steiner, L.~Fang, J.~Bai, and
  S.~Chintala, ``Pytorch: An imperative style, high-performance deep learning
  library,'' in \emph{Advances in Neural Information Processing Systems 32},
  H.~Wallach, H.~Larochelle, A.~Beygelzimer, F.~d\textquotesingle
  Alch\'{e}-Buc, E.~Fox, and R.~Garnett, Eds.\hskip 1em plus 0.5em minus
  0.4em\relax Curran Associates, Inc., 2019, pp. 8024--8035. [Online].
  Available:
  \url{http://papers.neurips.cc/paper/9015-pytorch-an-imperative-style-high-performance-deep-learning-library.pdf}
\BIBentrySTDinterwordspacing

\bibitem{chillotti2021programmable}
I.~Chillotti, M.~Joye, and P.~Paillier, ``Programmable bootstrapping enables
  efficient homomorphic inference of deep neural networks,'' in
  \emph{International Symposium on Cyber Security Cryptography and Machine
  Learning}.\hskip 1em plus 0.5em minus 0.4em\relax Springer, 2021, pp. 1--19.

\bibitem{lee2022privacy}
J.-W. Lee, H.~Kang, Y.~Lee, W.~Choi, J.~Eom, M.~Deryabin, E.~Lee, J.~Lee,
  D.~Yoo, Y.-S. Kim \emph{et~al.}, ``Privacy-preserving machine learning with
  fully homomorphic encryption for deep neural network,'' \emph{IEEE Access},
  vol.~10, pp. 30\,039--30\,054, 2022.

\end{thebibliography}

\end{document}